# Wavelet-Based Mel-Frequency Cepstral Coefficients for Speaker Identification using Hidden Markov Models

Mahmoud I. Abdalla and Hanaa S. Ali

**Abstract**—To improve the performance of speaker identification systems, an effective and robust method is proposed to extract speech features, capable of operating in noisy environment. Based on the time-frequency multi-resolution property of wavelet transform, the input speech signal is decomposed into various frequency channels. For capturing the characteristic of the signal, the Mel-Frequency Cepstral Coefficients (MFCCs) of the wavelet channels are calculated. Hidden Markov Models (HMMs) were used for the recognition stage as they give better recognition for the speaker's features than Dynamic Time Warping (DTW). Comparison of the proposed approach with the MFCCs conventional feature extraction method shows that the proposed method not only effectively reduces the influence of noise, but also improves recognition. A recognition rate of 99.3% was obtained using the proposed feature extraction technique compared to 98.7% using the MFCCs. When the test patterns were corrupted by additive white Gaussian noise with 20 dB S/N ratio, the recognition rate was 97.3% using the proposed method compared to 93.3% using the MFCCs.

## 1 INTRODUCTION

Recent data on mobile phone users all over the world, and the number of telephone landlines in operation confirm that voice is the most accessible biometric, as no extra acquisition device or transmission system is needed. This fact gives voice an advantage over other biometrics especially when remote users or systems are taken into account [1]. Speaker recognition is the process of automatically recognizing who is speaking based on individual information included in the speech waves. Any utterance contains information about the words being spoken and also about the identity of the speaker. In a speech recognition system we wish to select the first type of feature and ignore the second; in a speaker recognition system we wish to do just the opposite. The speaker's voice can be used to verify the identity of the person and allow accessing to services such as banking by telephone, database access services, security control for confidential information areas and remote access to computers [2]. Speaker recognition systems are classified as text-dependent and text-independent. Text-dependent systems require a user to pronounce some specified utterances containing the same text as the training data. Text-independent means that there is no limitation on the text used in the system.

The most important parts of a speaker recognition system are the feature extraction and the recognition methods. The feature extraction step converts the properties of the signal which are important for the pattern recognition task to a format that simplifies the distinction of the classes. The recognition step aims to estimate the general extension of the classes within feature space from a training set [2]. For feature extraction, Linear Predictive Coding (LPC) of speech has proved to be a valid way to compress the spectral envelope in an all-pole model [1]. Most speech recognition systems use the so-called Mel-Frequency Cepstral Coefficients (MFCCs) and its first and sometimes second derivative in time to better reflect dynamic changes [3]. For the recognition step, the problem of text-dependent speaker recognition is a problem of comparing a sequence of feature vectors to a model of the user. For this comparison, there are two methods that have been widely used, template based methods and statistical methods. Dynamic Time Warping (DTW) is one of the most widely used template based methods [1]. Statistical methods and in particular Hidden Markov Models (HMMs) tend to be used more often than template based methods. They provide more flexibility; allow using speech units from sub-phoneme units to words and enabling the design of text-prompted systems [4].

In this paper, a robust feature extraction algorithm for speech signals is proposed. This algorithm depends on combining both the wavelet transform and the MFCCs for the feature extraction stage. First, the wavelet transform is applied to decompose the speech signal into two different frequency channels. The components of the low frequency channel are the approximations, while the high frequency channel components are the details. The decomposition process can be iterated with successive approximations being decomposed. Second, for capturing the characteristics of the individual speakers, the MFCCs of the approximations and detail channels are calculated. Based on this mechanism, the multi-resolution features of the speech signal can easily be extracted using the wavelet decomposition and calculating the related coefficients. The proposed technique is used in the feature extraction stage of a text-dependent speaker identification system. HMMs are used for the identification stage and compared to the DTW template based models.

The rest of this paper is organized as follows. Section 2 gives a description of the proposed feature extraction technique and provides detailed description of each constituting part. Section 3 introduces the recognition techniques used. The experiments

- *M.I. Abdalla is with the Department of electronics and communications, Zagazig University, Egypt.*
- *H.S. Ali is with the Department of electronics and communications, Zagazig University, Egypt.*

and the results obtained are given in section 4. Concluding remarks are given in section 5.

## 2 Description of the Wavelet-Based MFCCs Feature Extraction Technique for Representing Speech Signals

Speech is a complicated signal produced as a result of several transformations occurring at several different levels: semantic, linguistic, articulatory and acoustic. Differences in these transformations appear as differences in the acoustic properties of the speech signal [5]. An important problem in speech recognition systems is to determine a representation that is well adapted for extracting information content of speech signals. In general, transforming a signal to a different domain is done to get a better representation of the signal. For recognition, better means having more ability to separate signals which belong to separate classes or categories in the new domain than in the original domain.

### 2.1 Wavelet Transform

Wavelet analysis is a relatively new mathematical discipline, which has generated much interest in both theoretical and applied mathematics over the past decade. Wavelets have the ability to analyze different parts of a signal at different scales. The wavelet transform (WT) is a transformation that provides time-frequency representation of the signal. The continuous one dimensional WT is a decomposition of $f(t)$ into a set of basis function $\Psi_{a,b}(t)$ called wavelets[6]:

$$w(a,b) = \int f(t)\psi^{*}{}_{a,b}(t)dt \qquad (1)$$

The wavelets are generated from a single mother wavelet $\Psi(t)$ by dilation and translation

$$\psi_{a,b}(t) = \frac{1}{\sqrt{a}}\psi\left(\frac{t-b}{a}\right) \qquad (2)$$

Where:
$f(t)$ is the signal to be analyzed, a is the scale, and b is the translation factor. $\Psi(t)$ is the transforming function and is called the mother wavelet. Filters of different cutoff frequencies are used to analyze the signal. A continuous wavelet transform can operate at every scale. Also the analyzing wavelet is shifted smoothly over the full domain of the analyzed function. In discrete wavelet transform (DWT), scales and positions of powers of two are chosen. An efficient way to implement this scheme using filters was developed by Mallat in 1988 [7]. Given a signal S of length N, the DWT consists of $\log_2 N$ stages at most. The first step produces, starting from S, two sets of coefficients: approximation coefficients CA1 and the detail coefficients CD1. These vectors are obtained by convolving S with a low pass filter for approximations, and with a high pass filter for details, followed by dyadic decimation. The next step splits the approximation coefficients CA1 into two parts using the same scheme, replacing S by CA1 and producing CA2 and CD2, and so on. This technique is most effective when it is applied to the detection of short-time phenomena, discontinuities, or abrupt changes in the signal.

### 2.2 Mel-Frequency Cepestral Coefficients (MFCCs)

In sound processing, the mel-frequency cepstrum (MFC) is a representation of the short-term power spectrum of a sound, based on a linear cosine transform of a log power spectrum on a nonlinear Mel scale of frequency. MFCCs are coefficients that collectively make up an MFC. The difference between the cepstrum and the mel-frequency cepstrum is that in the MFC, the frequency bands are equally spaced on the Mel scale, which approximates the human auditory system's response more closely than the linearly-spaced frequency bands used in the normal cepstrum. This frequency warping can allow for better representation of sound [8]. Fig. 1 illustrates extraction of MFCCs. The first step is to divide the speech signal into blocks using overlapping smooth windows such as Hamming, Hanning, etc. The next step is to take the Discrete Time Fourier Transform (DTFT) of the windowed signal. Next, the square of the DTFT of the windowed signal is calculated. The outputs of the fourth step are the mel-scaled filter bank energies. The fifth step involves calculating the logarithm of the mel-scaled filter bank energies. The last step involves taking the Discrete Cosine transform (DCT) of the mel-scaled log-filter bank energies to calculate MFCCs.

### 2.3 Wavelet-Based MFCCs Feature Extraction Technique

Speech signals contain two types of information, time and frequency. In time space, sharp variations in signal amplitude are generally the most meaningful features. In the frequency domain, although the dominant frequency channels of speech signals are located in the middle frequency region, different speakers may have different responses in all frequency regions [9]. Thus, the traditional methods which just consider fixed frequency channels may lose some useful information in the feature extraction process. In this paper, the multi-resolution decomposing technique using wavelet transform is used. Based on this technique, one can decompose the speech signal into different resolution levels. The characteristic of multiple frequency channels and any change in the smoothness of the signal can then be detected to perfectly represent the signals. Then, the MFCCs are applied to the wavelet channels to extract features characteristics. MFCCs as previously stated, has the advantage that they can represent sound signals in an efficient way because of the frequency warping property. In this way, the advantages of both techniques are combined in the proposed technique.

## 3 Recognition Techniques

In this section, a brief description of the recognition techniques used in this work is introduced. In speaker identification, the goal is to design a system that minimizes the probability of identification errors. Thus, the objective is to discriminate between the given speaker and all other speakers. This is done by computing a match score. This score is a measure of the similarity between the input feature vectors and some model. There are two types of models: template models and stochastic models. In template models, the pattern matching is deterministic. The align ment of the observed frames to template frames is selected to minimize a distance measure value.

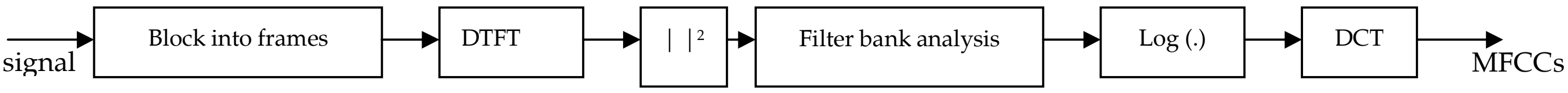

Fig. 1. Extraction of the MFCCs

In stochastic models, the pattern matching is probabilistic. The result is a measure of the likelihood, or conditional probability, of the observation given the model. [5]. DTW is one of the most important template modeling techniques and HMM is one of the most important stochastic modeling techniques.

### 3.1 Dynamic Time Warping (DTW)

DTW is a well known algorithm in many areas. This technique was first introduced in 60s and extensively explored in 70s by application to the speech recognition. The pattern matching in DTW technique is performed by seeking an optimal mapping from the test signal to the template signal, meanwhile allowing a nonlinear, monotonic distortion (warping) in the test signal [10],[11],[12].

The dynamic time warping problem is stated as follows: given two time series X and Y of lengths $|X|$ and $|Y|$, where

$$X = x_1, x_2, \ldots, x_i, \ldots, x_{|X|}$$
$$Y = y_1, y_2, \ldots, y_i, \ldots, y_{|Y|} \qquad (3)$$

A warp path W is constructed as follows

$$W = w_1, w_2, \ldots, w_K \qquad \max(|X|,|Y|) \le K \langle |X| + |Y| \qquad (4)$$

Where K is the warp path length and k$^{th}$ element of the warp path is

$$w_k = (i, j) \qquad (5)$$

Where i is an index from time series X, and j is an index from time series Y. The warp path must start at the beginning of each time series at w1= (1, 1) and finish at the end of both time series at $w_k$= (|X|, |Y|). In this way, every index of both time series is used in the warp path. There is also a constraint on the warp path that forces i and j to be monotonically increasing in the warp path. Every index of each time series must be used. Stated more formally:

$$w_k = (i, j), w_{k+1} = (i', j') \qquad i \le i' \le i+1, \qquad j \le j' \le j+1 \qquad (6)$$

The optimal path is the minimum distance warp path, where the distance of the warp path W is

$$Dist(W) = \sum_{k=1}^{k=K} Dist(w_{ki}, w_{kj}) \qquad (7)$$

Where, $Dist(W)$ is the Euclidean distance of the warp path W, and $Dist(w_{ki}, w_{kj})$ is the distance between the two data point indexes, one point from X and one from Y, in the k$^{th}$ element of the warp path [11].

### 3.2 Hidden Markov Models (HMMs)

HMMs are used to model a stochastic process defined by a set of states and transition probabilities between those states, where each state describes a stationary stochastic process and the transition from one state to another state describes how the process changes its characteristics in time. Each state of the HMM can model the generation or emission of the observed symbols or feature vectors using a stationary stochastic emission process. For a given observation (vector) however, the state in which the model has been when emitting that vector is not known. The underlying stochastic process is therefore hidden from the observer. Each state of the HMM will model a certain segment of the vector sequence of the utterance. These segments (which are assumed to be stationary) are described by the stationary emission processes assigned to the states, while the dynamic changes of the vector sequence will be modeled by transitions between the states of the HMM. In the states of the HMM, stationary emission processes are modeled, which are assumed to correspond with stationary segments of speech. Within those segments, the wide variability of the emitted vectors caused by the variability of the characteristic features of the speech signals should be allowed. In order to use a continuous observation density, some restrictions have to be placed on the form of the model probability density function (pdf) to insure that the parameters of the pdf can be reestimated in a consistent way. The most general representation of the pdf is a finite mixture of the form

$$b_j(O) = \sum_{m=1}^{M} c_{jm} \Pi(O, \mu_{jm}, U_{jm}) \qquad 1 \le j \le N \qquad (8)$$

Where:

$\Pi$ :any log-concave or elliptically symmetric density (e.g. Gaussian).

$c_{jm}$ : the mixture coefficient for the m$^{th}$ mixture in state j.

$\mu_{jm}$ : mean vector for the m$^{th}$ mixture component in state j.

$U_{jm}$ :covariance matrix for the m$^{th}$ mixture component in state j [13].

It is common to represent output distributions by Gaussian mixture densities. Mixture components can be considered to be a special form of sub-state in which the transition probabilities are the mixture weights [14]. The appropriate number of mixtures should be used and may be set individually for every state of the HMM. However, since a high number of mixtures also mean that a high number of parameters have to be estimated, some compromise between the granularity of the modeling and the reliability of the estimation of the parameters has to be found. To reduce the number of parameters and the computational effort simultaneously, it is often assumed that the individual features of the feature vector are not correlated. Then diagonal covariance matrices can be used instead of full covariance matrices. This reduces the number of parameters and the computational effort. As a consequence, a high number of mixtures can be used, and the correlation of the feature vectors can thus be modeled to a certain degree by the combination of diagonal mixtures [3].

### 3.2.1 Building HMMs

To estimate the parameters of the model, some observations sequences are used to adjust the model parameters to maximize P (O|λ) [13],[15]. In other words, derive the maximum likelihood estimate of the parameters of the HMM given a dataset of output sequences. No tractable algorithm is known for solving this problem exactly, but a local maximum likelihood can be derived efficiently using the Baum-Welch algorithm. The Baum-Welch algorithm is also known as the forward-backward algorithm, and is a special case of the Expectation-maximization algorithm. Given the observation sequences and the model, the optimal corresponding state sequence should be found. The problem is to find the state sequence that is most likely to have generated that output sequence. This requires finding a maximum over all possible state sequences. There is no exact and unique solution for this problem, but in practice, an optimality criterion is considered to solve the problem. There are several optimality criteria that can be applied. The Viterbi algorithm can be used to solve this problem efficiently.

Given the observation sequence $O=O_1\ O_2\ \ldots O_T$ and a model λ, compute P (O|λ), the probability of the observation sequence, given the model. This problem is evaluation or scoring problem. In case of trying to choose among several models, this solution give us the model which best matches the observation. This can be done efficiently using the forward algorithm. A detailed description of these algorithms can be found in [13].

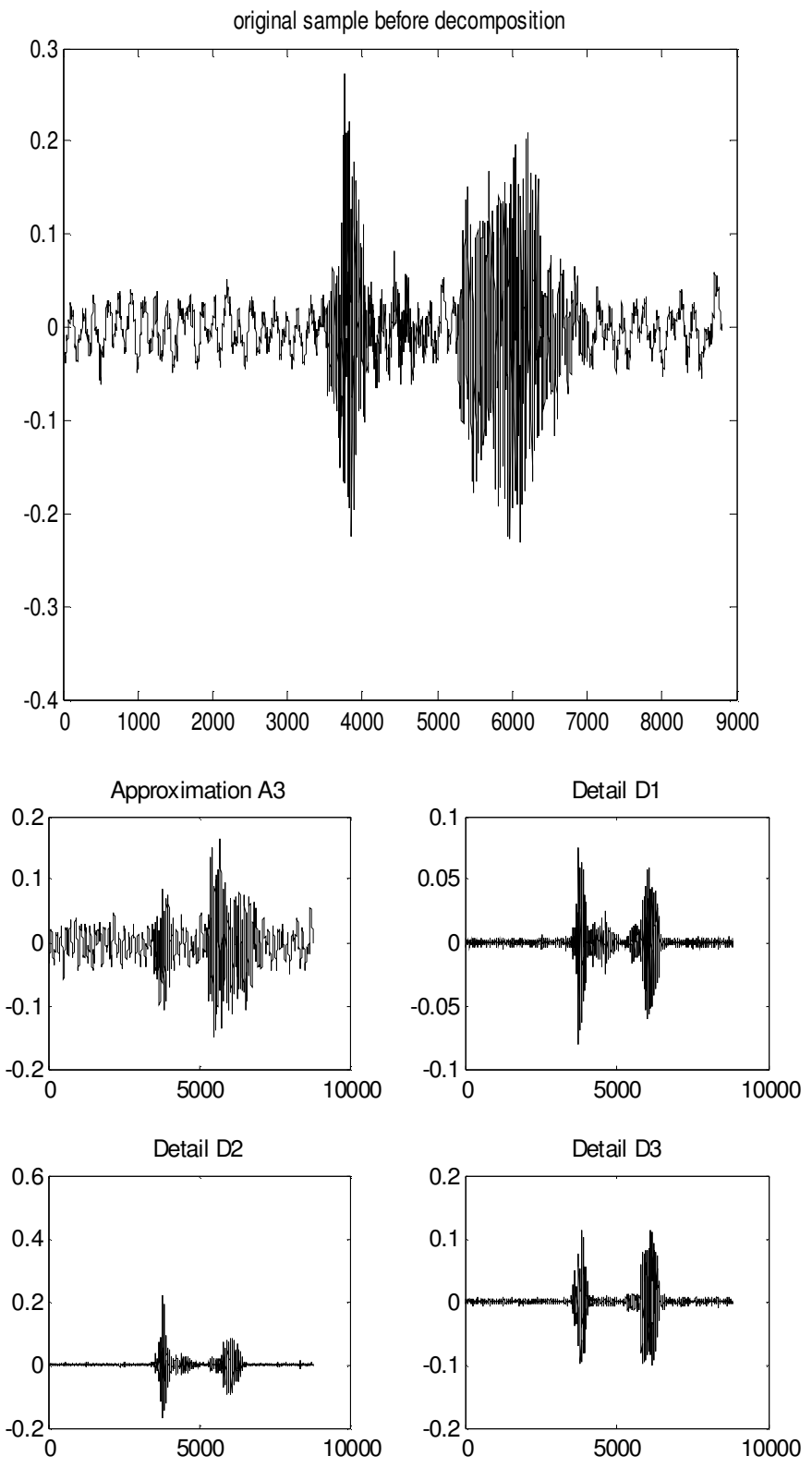

Fig. 2. (a) A sample speech signal, (b) Its wavelet channels

## 4. EXPERIMENTS AND RESULTS

The database contains the speech data files of 30 speakers. These speech files consist of isolated Arabic words. Each speaker repeats each word 15 times, 10 of the utterances are for training and 5 for testing. The data were recorded using a microphone, and all samples are stored in Microsoft wave format files with 8000 Hz sampling rate, 16 bit PCM and mono channels.

Using the proposed feature extraction technique, the wavelet transform is applied the voice signal, yielding approximations and details channels at different levels, and then the MFCCs are used to extract features from the wavelet channels. The number of extracted speech features is proportional to the number of decomposition levels. Although more decomposition processes can obtain more information from the input signals, the computational complexity and the number of useless features will increase greatly. In this work, the signals are decomposed at level 3 using db8 wavelet.Fig. 2 shows a sample speech signal for the first person and its wavelet channels. For the MFCCs, the Mel filter bank is designed with 20 frequency bands. In the calculation of all the features, the speech signal is partitioned into frames; the frame size of the analysis is 256 samples with 100 samples overlapping.

DTW is used for the recognition process. In this technique, a comparison is made between an input utterance template and the reference template. It uses an optimum time expansion/compression function for producing non-linear time alignment.. The results showed that a recognition rate of 98.6% is achieved using the wavelet based MFCCs features compared to 98% using MFCCs only. Fig. 3.a shows the DTW results comparing a sample from a speaker with another sample from the same speaker using the proposed technique, giving a distance of 23.1867 over the optimal path, while Fig. 3.b shows the results when comparing a sample from a speaker with a sample of a another speaker giving a distance of 151.8155 over the optimal path. The graphical output consists of three parts, the first part represents the matrix of local distances, the second represents the matrix of cumulated distances and the third is the DTW path, which tends to get darker blocks since it maximizes the matching performance. Note that minimizing the distance is equivalent to maximizing similarity. These results are obtained using Matlab.

Using DTW, a test vector sequence is aligned to each of the training sequences such that a certain distance measure (Euclidean in this work) is minimized. Therefore, the algorithm can handle variations about the length of the phonemes an utterance consists of. The traditional way of preparing the reference templates is by selecting one sample from each speaker and considering it as a reference template. It is noted from the experiments that using a single reference template has a disadvantage that it is not robust to the speech signal variability. This is because it is almost impossible for a person to repetitively speak a word exactly in the same way. The speech signal produced would vary according to many factors. So, multiple examples have to be prepared beforehand. All variability of speech that naturally occurs if a word is spoken at different times had to be reflected by the prototype vector se-quences. The number of prototypes that should be stored in is quite large. The feature vectors of the speakers should be represented in a generic form in order to recognize them. One would like to look for a model for each person instead of storing samples of his features. For these reasons, the second stage of the experiments was to use HMMs to model statistically the characteristic features of the phonemes that are present in the utterances.

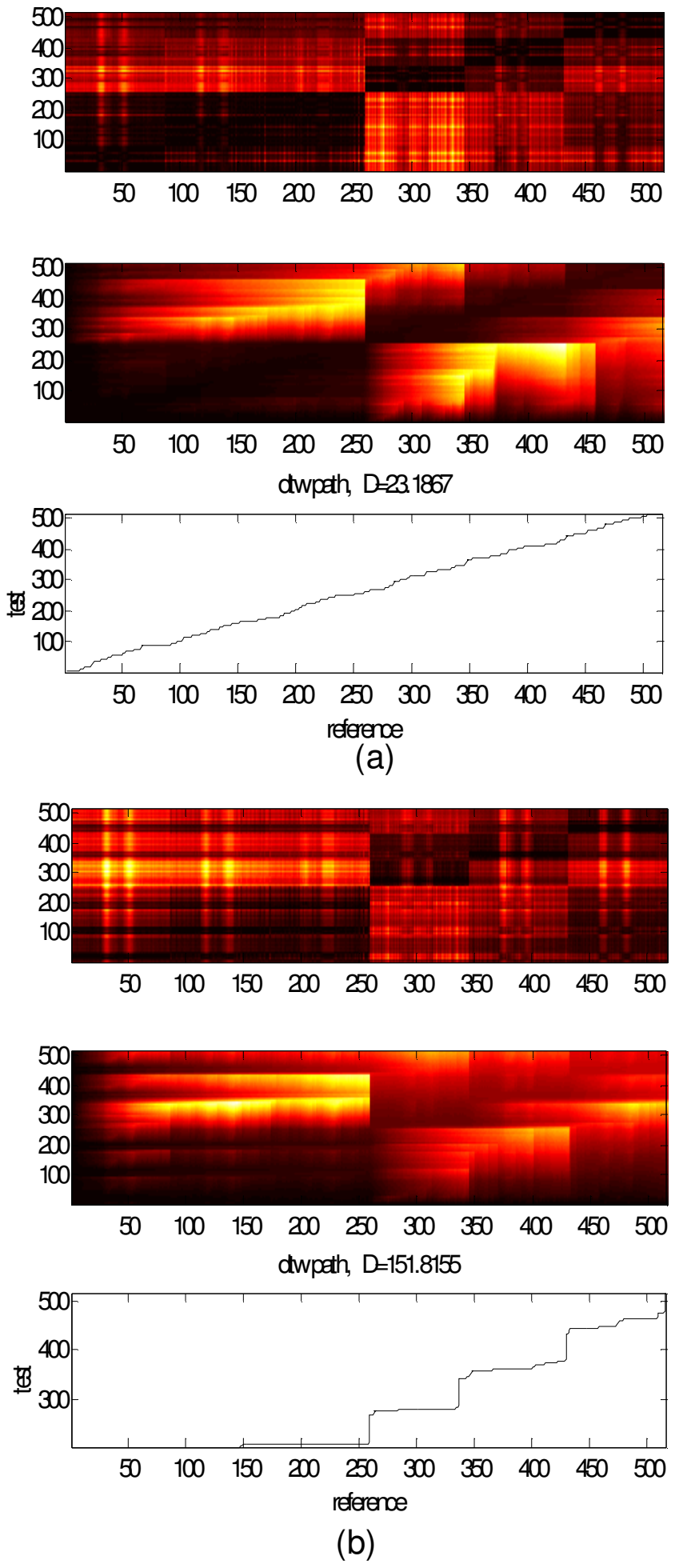

Fig. 3. Dynamic time warping results: (a) comparing a sample from a speaker with another sample from the same speaker, (b) comparing a sample from a speaker with a sample of a another speaker

A recognition system was developed using the Hidden Markov toolbox for use with Matlab, implementing a 4 states left-to-right transition model for each speaker, the probability distribution on each state was modeled as a 8 mixtures Gaussian with diagonal covariance matrix. HMMs are trained using the Baum-Welch algorithm [13]. This algorithm is used to derive the maximum likelihood estimate of the parameters of the HMMs given a dataset of output sequences using an iterative procedure. It works by guessing initial parameter values, then estimating the likelihood of the data using the current parameters. These likelihood values are then used to re-estimate the parameters iteratively until a local maximum is reached. In order to perform speaker identification using HMMs, a model is built for each person and the model parameters that optimize the likelihood of the training set observation vectors are estimated. For each unknown person who is to be recognized, features are extracted from his voice sample, followed by calculation of model likelihood of all models, and followed by the selection of the person whose model likelihood is highest. HMMs are used for recognition with the proposed technique used for feature extraction, and the results are compared to HMMs used for recognition with the MFCCs alone used for feature extraction. Also in order to evaluate the performance of the proposed method in a noisy environment, the test patterns for five utterances are corrupted by additive white Gaussian noise to the original signal so that the signal to noise ratio (SNR) is 20 dB. The results are summarized in Table 1.

TABLE 1

RECOGNITION RATES USING THE PROPOSED AND THE MFCCS TECHNIQUES IN BOTH CLEAN AND NOISY ENVIRONMENT

| Speech signal | Feature extraction technique | Recognition rate |
|---|---|---|
| Original clean signal | Wavelet-based MFCCs | 99.3% |
| | MFCCs | 98.7% |
| Noisy signal with S/N=20dB | Wavelet-based MFCCs | 97.3% |
| | MFCCs | 93.3% |

The results show that for clean speech, an identification rate of 99.3% is achieved by the proposed feature extraction technique, while the identification rate is 98.7% using the MFCCs. When the test patterns are corrupted with Gaussian noise, the performance of the system using MFCCs features is affected significantly by the added noise with an identification rate of 93.3%, while the proposed technique demonstrate much better noise robustness with a satisfactory identification rate of 97.3%.

## 5. CONCLUSION

In this paper, we presented an effective and robust feature extraction technique, for deployment with speaker identification systems. Based on the time-frequency analysis of the wavelet transform, approximations and details resolutions channels are obtained. The MFCCs of the wavelet channels are calculated for capturing the characteristics of the speech signals. Results showed that the proposed technique gives better performance than MFCCs features. In addition, this technique reduces the problem of noise and improves efficiently the recognition rate when dealing with noisy speech signals compared to the MFCCs which operate well only in clean environment. HMMs were used for the recognition stage as they give more advanced representation than DTW for the features of a certain speaker, the characteristics features for the phonemes of the utterances are modeled statistically using these models.

## ACKNOWLEDGEMENT

The authors would like to thank Professor Andrew Morris (Research Associate, Dept. of Phonetics, Saarbrücken University, Germany) for helpful discussion through emails.